\documentclass[journal]{IEEEtran}
\usepackage{ifpdf}
\usepackage{url}

\ifCLASSINFOpdf
\else
\fi

\usepackage{graphicx}
\usepackage{amssymb}
\usepackage[cmex10]{amsmath}
\begin{document}
\title{Dynamics of a micro-VCSEL operated in the threshold region under low-level optical feedback}
\author{Tao~Wang,
        Xianghu~Wang,
        Zhilei~Deng,
        Jiacheng~Sun,
        Gian~Piero~Puccioni,
        Gaofeng~Wang,
        and Gian~Luca~Lippi
\thanks{T. Wang, X. ~Wang, Z. ~Deng, J. ~Sun and G. Wang are with School of Electronics and Information, Hangzhou Dianzi University, Hangzhou, 310018, China, e-mail: wangtao@hdu.edu.cn}
\thanks{G. P. Puccioni is with Istituto dei Sistemi Complessi, CNR, Via Madonna del Piano 10, I-50019 Sesto Fiorentino, Italy, e-mail: gianpiero.puccioni@isc.cnr.it}
\thanks{G. L. Lippi is with Universit\'e C\^ote d'Azur, Institut de Physique de Nice (INPHYNI), UMR 7010 CNRS, 1361 Route des Lucioles, F-06560 Valbonne, France, e-mail: Gian-Luca.Lippi@inphyni.cnrs.fr}}

\maketitle

\begin{abstract}
Semiconductor lasers are notoriously sensitive to optical feedback, and their dynamics and coherence can be significantly modified through optical reinjection. We concentrate on the dynamical properties of a very small (i.e., microscale) Vertical Cavity Surface Emitting Laser (VCSEL) operated in the low coherence region between the emission of (partially) coherent pulses and ending below the accepted macroscopic lasing threshold, with the double objective of:  1. studying the feedback influence in a regime of very low energy consumption; 2. using the micro-VCSEL as a surrogate for nanolasers, where measurements can only be based on photon statistics.
The experimental investigation is based on time traces and radiofrequency spectra (common for macroscale devices) and correlation functions (required at the nanoscale).  Comparison of these results confirms the ability of correlation functions to satisfactorily characterize the action of feedback on the laser dynamics.  Numerical predictions obtained from a previously developed, fully stochastic modeling technique provide very close agreement with the experimental observations, thus supporting the possible extension of our observations to the nanoscale.
\end{abstract}


\IEEEpeerreviewmaketitle

\section{Introduction and objectives}

Optical communications have been the prime mover behind the investigation of optical feedback on the emission properties of semiconductor lasers, due to the extreme sensitivity of edge-emitting devices to even very low reinjection levels (e.g., from fiber ends~\cite{Henry1986}).  The noise and coherence properties, as well as the emitter's dynamical stability, are thus modified, with detailed features depending on whether the reinjected light fraction carries phase information~\cite{Serrat2003,Ohtsubo2017} (coherence length $L_c > L_f$, $L_f$ feedback length) or whether only the reinjected photon fraction counts~\cite{Cohen1990,Ju2005} ($L_c < L_f$).  The large number of investigations dedicated to the study of feedback is reviewed in different papers, depending on the scope:  for instance, optical transmission systems~\cite{Petermann1995} or understanding the dynamical features of semiconductor lasers with optical reinjection~\cite{Panajotov2011}.

Developed in the 1980's and 1990's, Vertical Cavity Surface Emitting Lasers (VCSELs) have become the most widespread coherent light sources thanks to their versatility, good beam quality, ease of integration and single-longitudinal mode emission.  In addition, they are less sensitive to optical feedback than their edge-emitting counterparts, rendering them more attractive in many applications, in spite of their polarization sensitivity on which have focussed most of the optical reinjection investigations (cf., e.g.,~\cite{Giudici1999,Hsu2001,Sondermann2003s,Sciamanna2003}).  Our work focuses on the basic understanding of operation regimes of very small devices which, in the future, could be used for transmissions and interconnects (e.g. datacenter applications~\cite{Tatum2015,Kuchta2015}).  Thus, we look at the dynamics of VCSELs due to non-polarization-selective feedback~\cite{Besnard1999,Naumenko2003,Sondermann2003}.  An overview of the phenomenology observed in VCSELs with optical feedback can be found in~\cite{Panajotov2012}.

VCSELs themselves are at the origin of laser miniaturization with the first design of a vertical semicounductor cavity~\cite{Soda1979,Iga2000}, which later branched along several independent directions (e.g., photonic-crystal-based devices~\cite{Hostein2010}).  The very low threshold and low power dissipation typical of nanolasers promise breakthroughs in a broad palette of applications~\cite{Ma2019}, which, for our purposes, cover light sources for all-optical chips~\cite{Smit2012}, data centers~\cite{Ghiasi2015,Tatum2015,Kuchta2015} and quantum information~\cite{Santori2002,Ates2009,Dousse2010,Schlehahn2016}.  However, their extremely small photon numbers render an in-depth characterization quite challenging.  Thus, aside from a couple of older investigations~\cite{Albert2011,Hopfmann2013} (even at the single-photon level~\cite{Carmele2013}), only recently have concerted efforts surfaced, covering optoelectronic feedback~\cite{Munnelly2017} or external light feedback both in microcavities~\cite{Holzinger2018a,Holzinger2018b,Holzinger2019} and in photonic-crystal-based Fano-lasers~\cite{Rasmussen2018a,Rasmussen2018b}.

Aiming at future on chip and datacenter applications, which require extremely low power consumption, we concentrate on the investigation of feedback on the emission at bias current levels in the region between the first light emission and the traditional threshold~\cite{Wang2015}, where we have shown that it is possible to obtain reliable pulse generation at lower energetic costs~\cite{Wang2019}.  Our objective is threefold:  understanding the dynamical influence of feedback on a small (mesoscale) laser at ultra-low bias (i.e., below the traditional threshold) where the deterministic dynamics mixes with stochastic behaviour~\cite{Wang2018C}; testing the predictive capabilities of fully stochastic modeling~\cite{Puccioni2015}; and probing the second-order (zero-delay) autocorrelation as a suitable dynamical indicator.  This last point aims at validating, on a micro-VCSEL~\cite{Wang2018S}, the use of the only technique currently usable to characterize nanolasers, the target devices for datacom applications.  

The need for photon statistics to investigate nanolaser behaviour (including their threshold properties) comes from the lack of available detectors with sufficient sensitivity and, simultaneously, electrical bandwidth to obtain a full characterization of the laser output.  The extremely low photon flux (typically at the sub-nW, or even pW level) renders interferometry challenging for routine tests.  Thus, photon statistics becomes the obvious choice, given that it possesses both the necessary time response and optical sensitivity.  However, since it only provides statistical information, the identification of the emission features must be based on models, or -- as is the case with our present proposal -- on the comparison with other techniques.  A discussion on the 
pertinency of microlaser investigations to learn about nanolaser behaviour and comparison between linear measurements and photon statistics can be found in~\cite{Wang2018S}.

\section{Experimental details and observations}

\begin{figure}[!t]
\centering
  \includegraphics[width=2.8in]{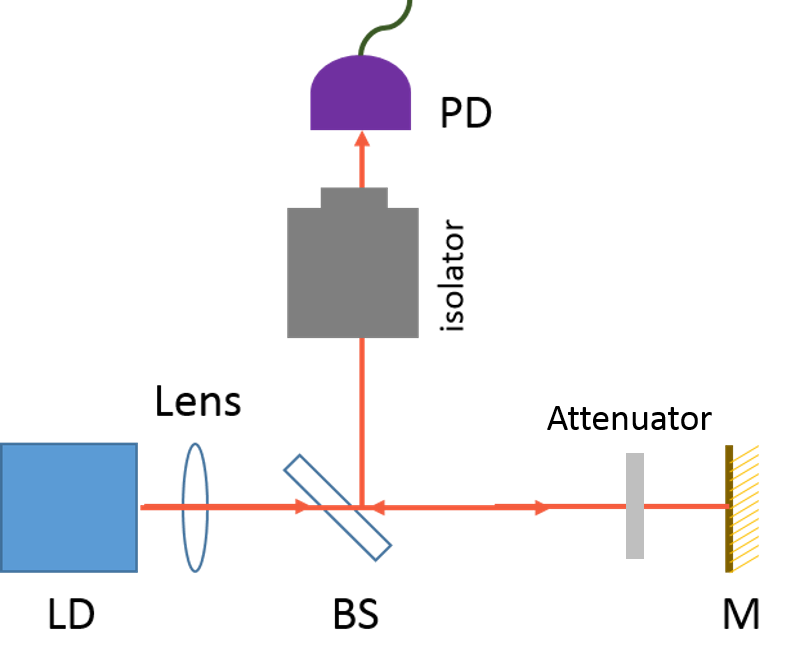}
  \caption{Schematic illustration of the experimental design: LD, semiconductor laser diode (VCSEL); BS, beam splitter; M, dielectric mirror; PD, fast photodetector. The semiconductor micro-VCSEL is temperature stabilized at $25^{\circ}$C, powered by a commercial dc power supply (Thorlabs LDC200VCSEL) with resolution $1\mu \rm A$ and accuracy $\pm 20 \mu \rm A$.}
  \label{EXP-setup}
\end{figure}

The experimental setup is shown in Fig.~\ref{EXP-setup}.  The laser is a VCSEL 980 (Thorlabs), designed for LAN data transmission at 2.5 Gb/s~\cite{Manufac}, electrically supplied by a stabilized current source (Thorlabs LDC200VCSEL), temperature-stabilized by a home-made controller to better than 0.1$^{\circ} C$, with estimated $\beta \approx 10^{-4}$~\cite{Wang2015}.  The nominal threshold current declared by the Manufacturer~\cite{Manufac} is typically $i_{th} \approx 2.2 \rm mA$, but can be as high as $i_{th,max} = 3.0 \rm mA$ for some devices, while the maximum operating current is $i_{max} = \rm 10 mA$ with corresponding maximum laser output $P_{max} \approx 1.85 \rm mW$.  The laser emits on a single polarization until a pump current value $i \approx 2.5 \rm mA$ (corresponding to the limit of the range we investigate) with a rejection ratio of approximately 24 dB (spontaneous emission is, of course, isotropic).  Additional technical information can be found in the Supplementary Information section of~\cite{Wang2015}.  

Comparison between the coherence properties of this device~\cite{Wang2015} and the manufacturer's specifications suggest a consistency between the maximum specified threshold current value and the threshold for laser coherence, since the latter is attained close to $i_{th,max}$~\cite{Wang2015}.  However lasing emission, in the form of partially coherent spikes, can already be obtained from $i \approx 1.26 \rm mA$~\cite{Wang2015}, thus providing an interesting pumping range below the coherence threshold where the influence of feedback on coherence buildup can be tested, and opening potential new windows for applications of this self-pulsing regime.

A dielectric, high-reflectivity mirror is aligned in front of the VCSEL at a distance $L_{ec}$ ($2 L_{ec} = (70 \pm 4)\, \rm cm$ -- i.e., an external cavity mode spacing $\Delta \nu_{ec} \approx 0.23 \rm GHz$) to provide the external cavity; inside the cavity we find a commercial, AR-coated collimator (identified as ``lens'' in Fig.~\ref{EXP-setup}), a non-polarizing beamsplitter (BS) to send part of the output to a detector, and a variable attenuator, used to fix the amount of power reinjected into the laser by the external cavity:  throughout the experiment we used a reinjection coefficient $\approx 1.5\%$.  This choice is made to match parasitic reinjection values which  would originate from residual and unwanted backreflections.  Since it is not a critical parameter (no dramatic changes appear as long as the reflectivity is kept low) and since the actual value would depend on the configuration (whether in free-space or integrated optics, etc.), we keep it constant throughout the experiment.

A Faraday isolator (QIOPTIQ8450-103-600-4-FI-980-SC, 40 dB isolation ratio) is placed in front of the fast, preamplified photodetector (Thorlabs PDA8GS, $9.5 \rm GHz$ bandwidth), to avoid backreflected contributions coming from the latter.
The electrical signal from the photodetector is digitized by a LeCroy Wave Master 8600A oscilloscope ($6 GHz$ analog bandwidth -- acquiring $1 \times 10^6$ points in all measurements). The data are stored in a computer through a GPIB interface controlled in Python.  The second-order autocorrelation function is numerically computed from the data trace acquired by the linear detector, as in~\cite{Wang2015}.

\begin{figure}[!t]
\centering
  \includegraphics[width=3.0in]{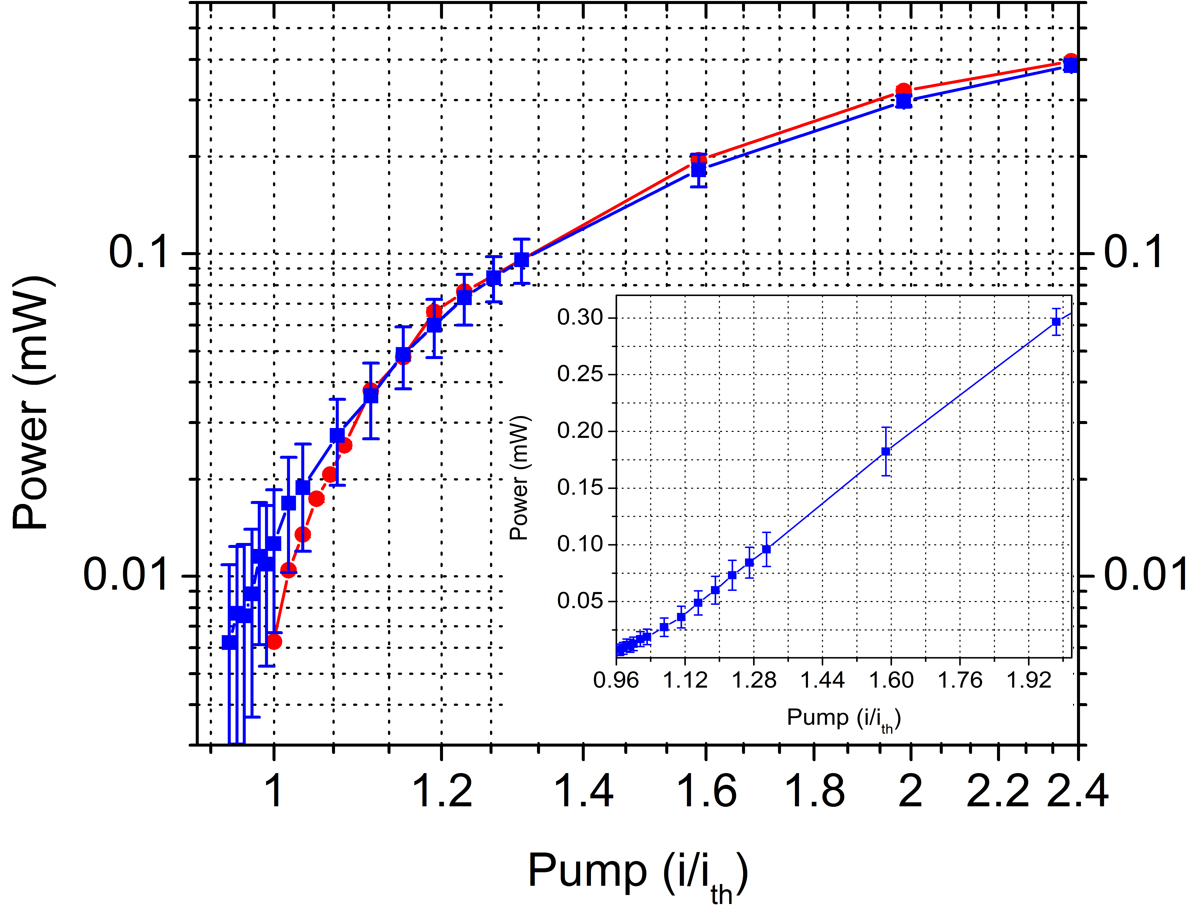}
  \caption{Main panel:  log-log laser response with (squares, blue online) and without feedback (circles, red online).  Inset:  detail in linear scale.}
  \label{Exp-averageintensity2}
\end{figure}

Fig.~\ref{Exp-averageintensity2} shows, in double-logarithmic scale, the average laser output {\it in the presence} of the external cavity (squares) for different values of the pumping current, compared to the same response in the absence of feedback (circles).  The error bars are computed from the fluctuations in the measured signals.  {\it In the absence} of feedback, emission from the laser in the form of irregular bursts has been observed for $i = 1.26 \rm mA$~\cite{Wang2015}, thus we choose this bias point as reference ($i_{th} = \rm 1.26 mA$ hereafter) to normalize all currents to a common value.  As in macroscopic lasers ($\beta \lessapprox 10^{-5}$), the addition of feedback introduces a reduction in the observed threshold ($\approx 4\%$, compatible with the findings of~\cite{Jiang1994}).  
The inset (Fig.~\ref{Exp-averageintensity2}) displays a detail of the laser response in linear scale, confirming that feedback lowers the emission threshold and proving that the largest fluctuations occur at $i \approx 1.6 i_{th}$.

\begin{figure}[!t]
\centering
  \includegraphics[width=3.0in]{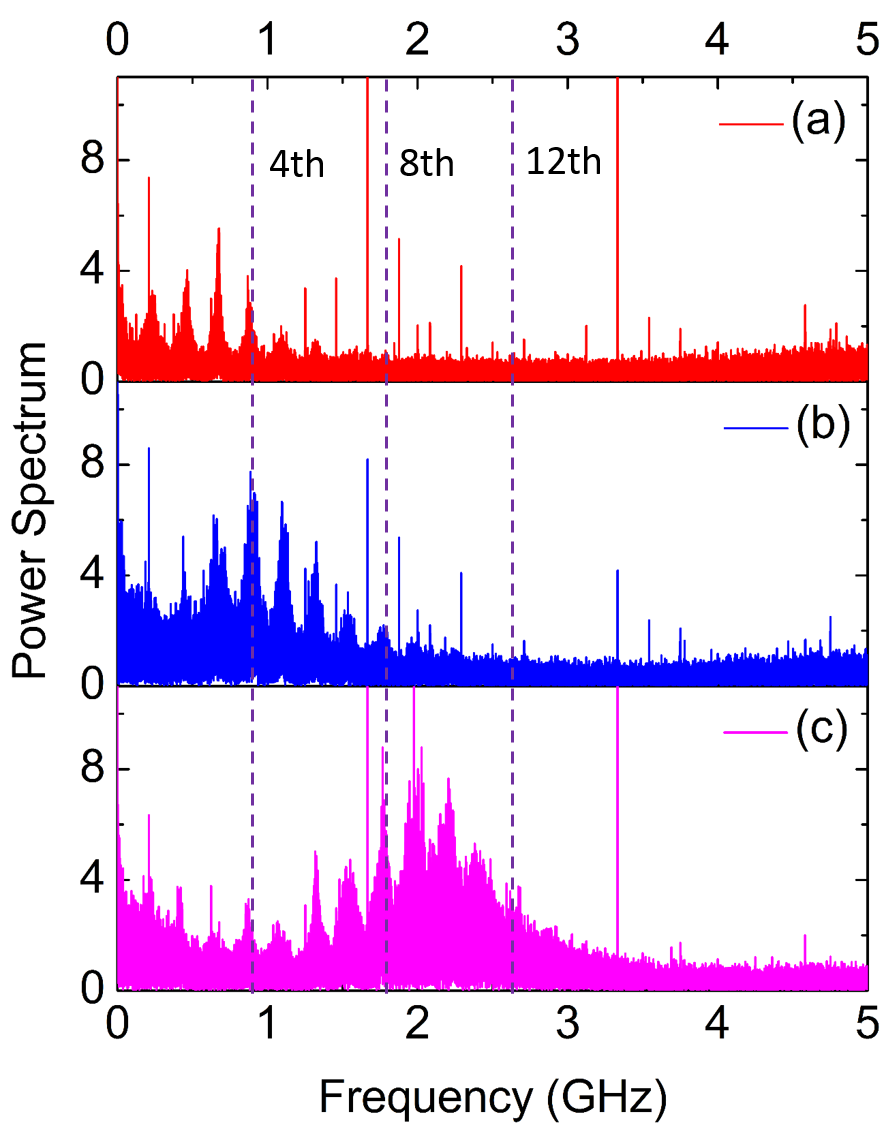}
  \caption{Radiofrequency spectra for (a) $i_{pump} = 1.00i_{th}$, (b) $1.10i_{th}$, and (c)$2.00i_{th}$.  The very narrow lines appearing in all spectra are artefacts due to spurious rf components presents in the environment.  The three dashed vertical lines, across the three panels, mark the position of the 4$^{\rm th}$, 8$^{\rm th}$, and 12$^{\rm th}$ harmonic component of the external cavity's roundtrip frequency.}
  \label{ESpec}
\end{figure}

Radiofrequency (rf) power spectra characterize the observed dynamics.  Fig~\ref{ESpec}(a) shows the rf spectrum at $i = i_{th}$:  peaks appear at the external cavity repetition frequency ($\Delta \nu_{ec} \approx 0.23 \, \rm GHz$) and its harmonics, where the third one, located close to an intrinsic ``resonance'' in the spiking (cf.~\cite{Wang2015,Wang2017,Wang2018} for its interpretation), presents a somewhat stronger amplitude and slightly narrower peak, but no additional feature (as in the free-running case~\cite{Wang2017}).

The onset of coherence is signalled by the appearance of a broad resonance (Fig.~\ref{ESpec}b) -- as in the free-running laser~\cite{Wang2016} -- onto which a comb of external cavity modes is superimposed.  A partial loss of ``long-term'' memory is signalled by drowning of the first two harmonic components of $\Delta \nu_{ec}$ into the low-frequency spectral enhancement ($0 \le \nu \lessapprox 0.25 \, \rm GHz$), compatible with weak LFFs.  Finally, for well-established laser operation (Fig.~\ref{ESpec}c) a stronger broad peak appears centered at $\nu \approx 2 \, \rm GHz$, signature of free-running laser operation~\cite{Wang2016}, upon which the same external cavity frequency comb is superposed.  It is interesting to notice, however, that the frequency harmonics of $\Delta \nu_{ec}$ are extended:  hints of the 12$^{\rm th}$ and 13$^{\rm th}$ component are visible here, while in panel (b) one could barely distinguish the 9$^{\rm th}$ component.  Also on the low-frequency side there is an improvement in the visibility of the harmonics, with even an identifiable trace of the 1$^{\rm st}$ component.  This suggests that the laser output is gaining in overall phase coherence by being capable of maintaining and partially imprinting the (relatively) long timescale features of the external cavity onto the laser output, while at the onset of coherent output~\cite{Wang2018} memory was very limited (panel b).  Notice that the low-frequency part of the spectrum, $\nu < 0.25 \, \rm GHz$, remains practically unchanged between panels (b) and (c) -- unlike the free-running laser~\cite{Wang2016} --, suggesting that the low-frequency part of the spectrum be a mixture of ``slow'' intrinsic dynamics and of LFFs, where the latter gain in relevance with bias (presumably due to growing coherence).

\begin{figure}[!t]
\centering
  \includegraphics[width=3.3in]{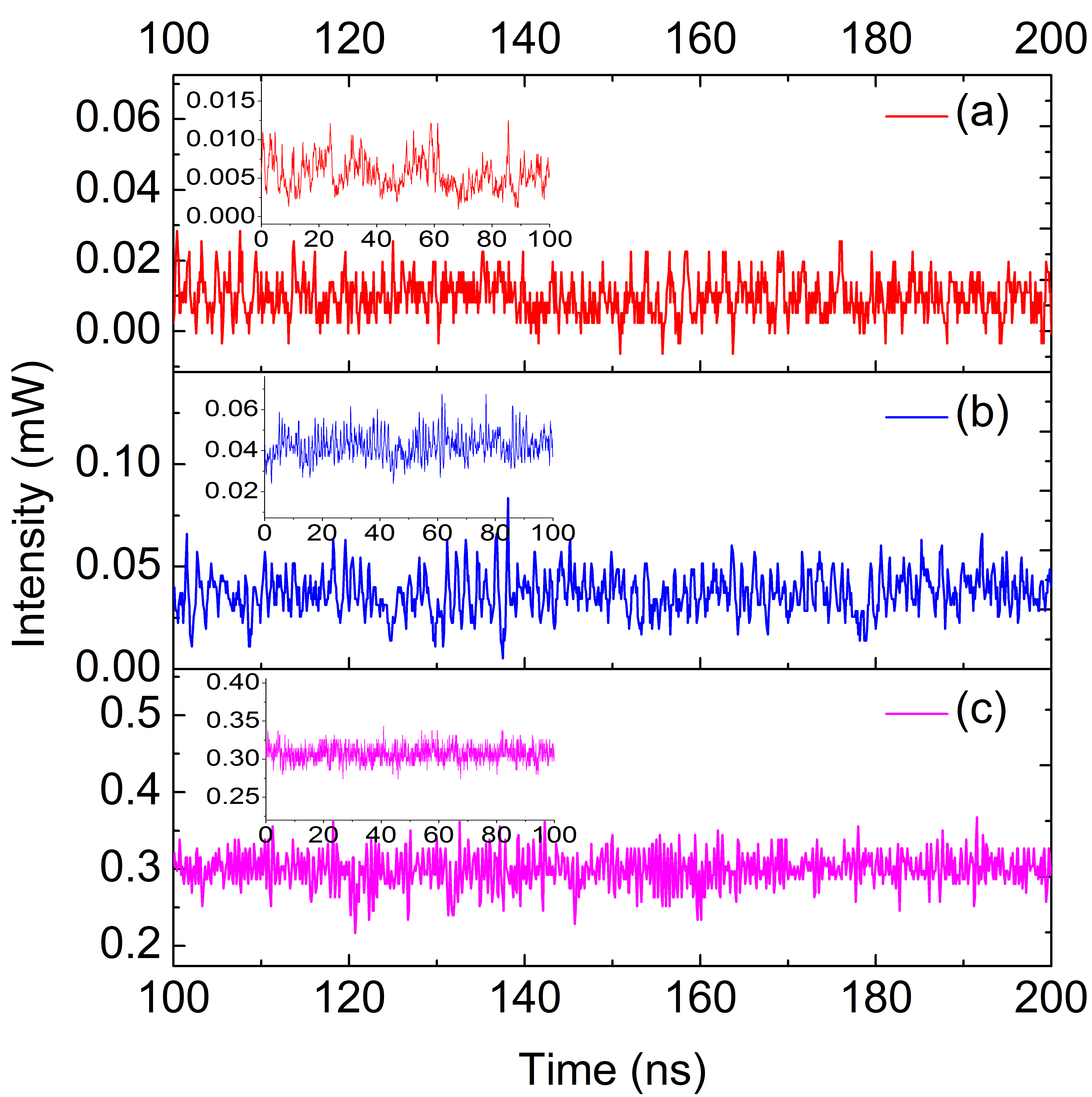}
  \caption{Temporal intensity data traces at (a)$1.00i_{th}$, (b)$1.10i_{th}$, and (c)$2.00i_{th}$.  The insets show corresponding sample traces in the free-running laser.}
  \label{EDynamics}
\end{figure}

Inspection of the temporal traces provides additional information.  At low pump rate (panel (a), Fig.~\ref{EDynamics}) the temporal output is rendered more regular by feedback (main panel) which {\it encourages} photon emission through re-injection (compare to the irregular bursts of the free-running regime -- inset).  A slightly negative residual bias in the detector's output produces, at times, ``negative'' values for output power (rather than drops to zero).  Larger fluctuations (i.e., involving sharper drops in photon emission) are instead observed at $i = 1.1 i_{th}$ (Fig.~\ref{EDynamics}(b)), but no strong differences are identifiable when comparing to the feedback-less case.  
Power drops (LFF) become more noticeable at larger pump values (Fig.~\ref{EDynamics}c) with dynamics clearly different from the free-running regime (inset).

The tools employed so far for the characterization of the experimental observations are those typical of macroscopic lasers.  We now consider indicators which become indispensable for micro- and nanoscale devices~\cite{Albert2011,Holzinger2018a,Holzinger2019}, but have only occasionally been used for macroscopic lasers~\cite{Sondermann2005}.  Photon counting is the only experimentally available technique with sufficient bandwidth and sensitivity for the very low photon fluence values of nanolasers.  It is based on the measurement of the second-order correlation function $g^{(2)}(\tau)$:
\begin{equation}
\label{defg2}
g^{(2)}(\tau) = \frac{\langle I(t+\tau)I(t) \rangle}{\langle I(t+\tau) \rangle \langle I(t) \rangle} \, ,
\end{equation}
where $I$ is the measured instantaneous laser intensity, $\tau$ is the time delay, and $\langle \cdot \rangle$ denotes statistical averaging. If the emitted light is coherent, $g^{(2)}(\tau) = 1$ independently of the $\tau$ value, while thermal radiation is characterized by $g^{(2)}(\tau=0) = 2$, with a gradual decay towards 1 as the emitted field acquires a degree of coherence.  Since our setup measures the dynamics of the intensity, $I$, with a temporal resolution $\Delta t_o = 0.1 \rm ns$, we can directly compute the correlation, eq.~(\ref{defg2}), from the time series~\cite{Wang2015,Wang2018}.   Notice that the filtering introduced by the limited electronic bandpass reduces the absolute values of the correlations~\cite{Wang2015,Wang2017A}, while maintaining its functional shape.

\begin{figure}[!t]
\centering
  \includegraphics[width=2.8in]{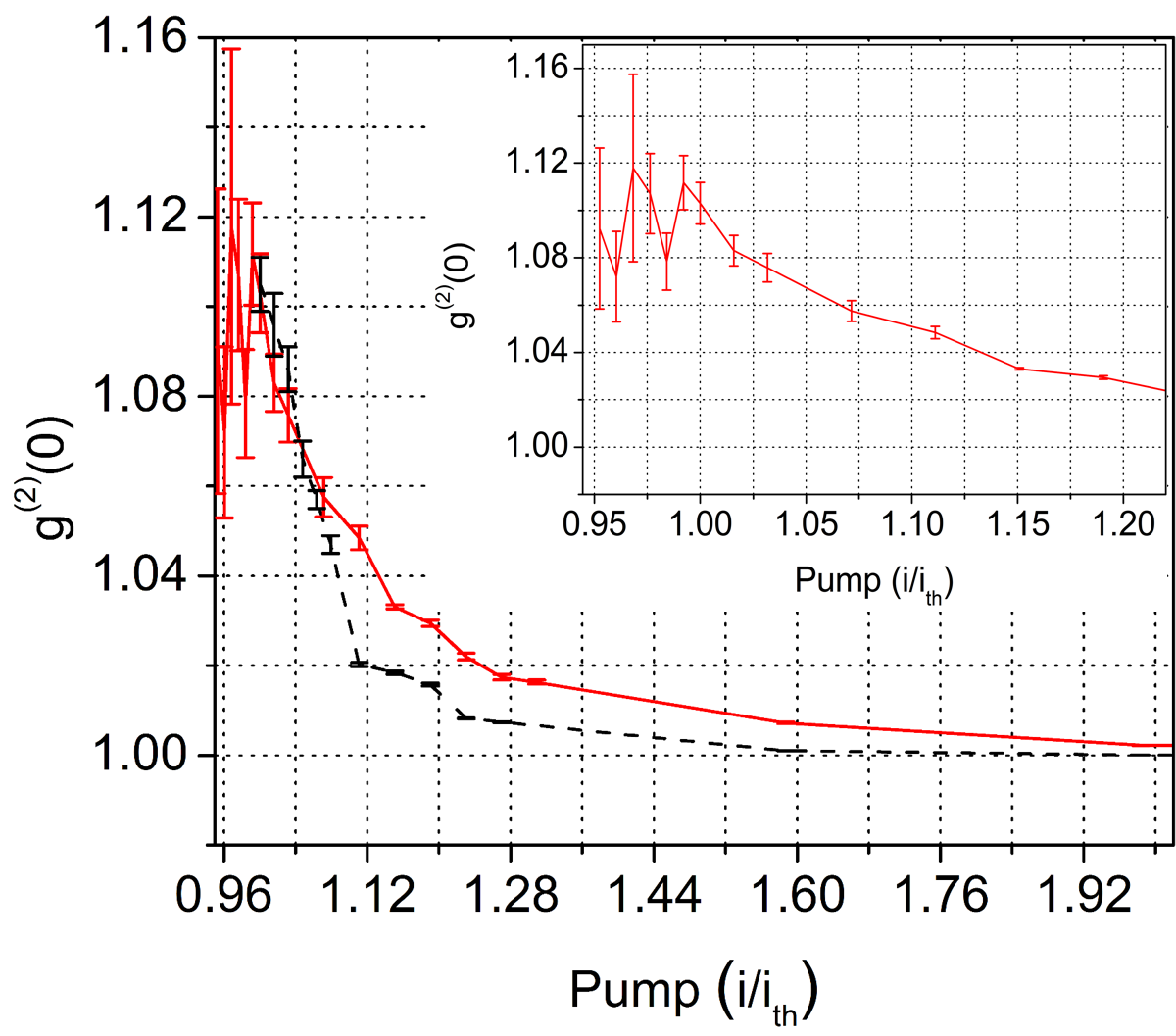}
  \caption{Measured second order autocorrelation ($g^{(2)}(\tau = 0)$) as a function of bias:  free-running laser (dashed, black, line);  fed-back laser (solid, red, line). Inset: detail of the autocorrelation in the presence of feedback for low pump values.
  }
  \label{Exp-autoenlarge3}
\end{figure}

Fig.~\ref{Exp-autoenlarge3} compares the zero-delay ($\tau = 0$) autocorrelation function obtained from data {\it in the absence} of feedback~\cite{Wang2015} (dashed, black line) and {\it in the presence} of feedback (solid, red line).  Below $i_{th}$ (cf. inset for details) feedback induces strong changes in the autocorrelation as shown not only by its ``oscillations'' in the mean value but also by the large error bars.  This regime ($i < i_{th}$) corresponds to the feedback-induced extension of the emission below threshold, where the photons reinjected into the laser by the external cavity act as an additional source of  ``spontaneous emission noise'', whence the strongly fluctuating correlations.  The autocorrelation is (slightly) lower in the presence of feedback than in its absence, in agreement with the temporal traces (Fig.~\ref{EDynamics}a) which show a less irregular signal with optical reinjection.  A more important observation, instead, is that coherence is degraded by feedback for $i > 1.04 i_{th}$.  Although this behaviour requires further investigation, it is plausible to relate its occurrence to the fact that even when the laser emits a (noisy) cw field ($i \gtrapprox 1.12 i_{th}$~\cite{Wang2018}), its coherence remains low, as proven by the lack of convergence of $g^{(2)}(0)$ towards 1 (reached only for  $i > 3 mA$~\cite{Wang2015}).  It appears that in this {\it less-coherent} regime optical reinjection reduces the, already low, coherence rather than enhancing it, as observable in larger devices and well above threshold.

\begin{figure}[!t]
\centering
  \includegraphics[width=3.3in]{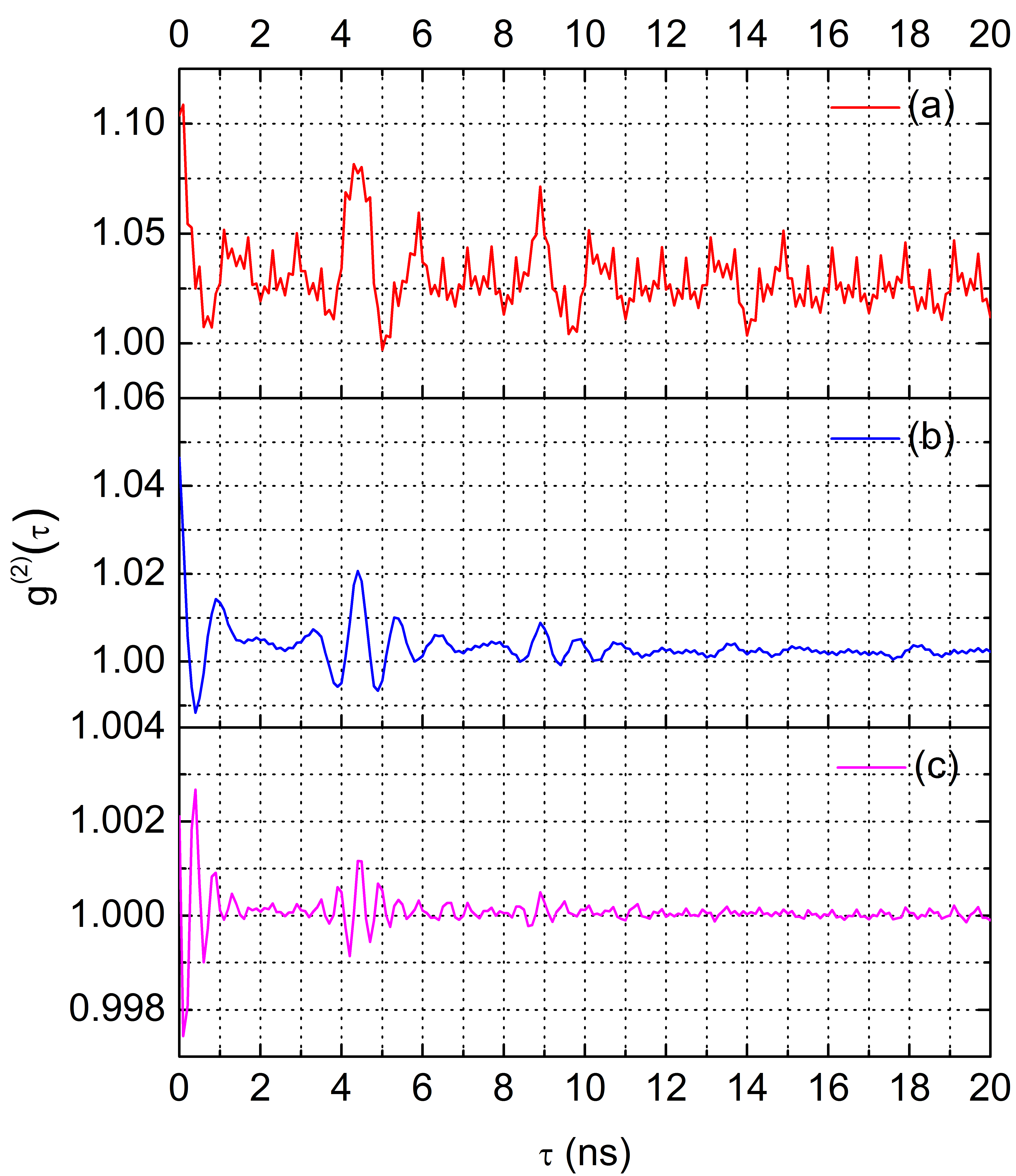}
  \caption{Time-delayed second-order autocorrelation for:  $i_{pump} = 1.00 \, i_{th}$ (a), $1.10 \, i_{th}$ (b) and $2.00\,  i_{th}$ (c).}
  \label{ECorrelation}
\end{figure}

The delayed ($\tau \ne 0$) autocorrelation, eq.~(\ref{defg2}), provides additional information on the laser dynamics for the three selected pump values already shown in Figs.~\ref{ESpec} and \ref{EDynamics}.  Fig.~\ref{ECorrelation}a shows a coherence revival at $\tau \approx 4.3 \rm ns$ with a second order multiple (the third is barely recognizable) which corresponds to the roundtrip time in the external cavity.  In the finer details, one can recognize the signature of an intrinsic periodicity with $\delta t \approx 0.6 \rm ns$ stemming from pulse repetition~\cite{Wang2015,Wang2017}.  However, since no precise spectral feature is recognizable in the self-spiking regime in the absence of feedback~\cite{Wang2015,Wang2017}, this component would be expected to be very small.  Alternately, the additional ripples superposed on the correlation signal (Fig.~\ref{ECorrelation}a) may also originate from the details of the spiking dynamics which, for the moment, remain unclear.

A drastic change in the laser output characteristics is visible in the autocorrelation at $i = 1.1 i_{th}$ (panel b) where the trace is much smoother and, in addition to the signature of the external cavity (as in panel a), one recognizes in the coherence revival two features:  1. an oscillation with period $T_{las} \approx 1 \rm ns$, modulating the revival peak and originating from the intrinsic laser resonance (spectral component at $\approx 1\, \rm GHz$ in Fig.~\ref{ESpec}b); and 2. the revival's width (two full oscillations in $g^{(2)}(\tau)$) which corresponds to the spectral width of the laser resonance ($\approx 2\, \rm GHz$ in Fig.~\ref{ESpec}b) -- broader than in the self-spiking regime.  The same holds at the larger pump value (Fig.~\ref{ECorrelation}c), where the ``faster'' oscillation in $g^{(2)}(\tau)$ comes from the larger central frequency value in Fig.~\ref{ESpec}c, with very similar spectral width.

The comparison offered by our experiment, capable of providing at the same time direct (through temporal traces and easy access to rf power spectra) and statistical information (through autocorrelations) enables us to compare the performance of the two techniques and confirm that correlations can be used to interpret experimental results obtained from much smaller lasers (e.g.,~\cite{Holzinger2019}), where only statistical measurements are possible.  Although experimental proof for our claim is currently missing, as we are validating the comparison between dynamical and statistical techniques for the first time, it is plausible that the concurrence of the information should hold irrespective of the laser size (i.e., cavity $\beta$), or degree of coherence of the emitted radiation.  Indeed, while at larger values of $\beta$ noise plays a more substantial role, correlation functions, spectra and time traces carry the corresponding information, even if its relevance may be at times more difficult to identify in one of the chosen indicators.  In addition, fully Stochastic Simulations~\cite{Puccioni2015} have been shown to well-reproduce  experimental observations~\cite{Wang2015} and can span the whole range of meso- and nanolaser scales~\cite{Wang2018S} (cf. Supplementary Material for additional information).  Below, we test this numerical technique to compare its predictions to our observations in the limit of incoherent feedback.

In addition to the limitations already found in matching experimental observations in ``large'' VCSELs~\cite{Torcini2006} to Lang-Kobayashi models~\cite{Lang1980}, theoretical support for experiments is difficult to obtain for small lasers, where the traditional rate-equations-based models (or even Maxwell-Bloch ones) cannot be used in conjunction with noise sources.  Although sufficiently far from threshold predictions of stochastic rate equations hold for macroscopic lasers, at the meso- (and nano-) scale there appear intrinsic violations of the necessary conditions for a proper Langevin noise description~\cite{Lippi2018,Lippi2019}, due to memory effects in the carriers typical of Class B lasers~\cite{Arecchi1984,Tredicce1985}.  We therefore resort to using a fully stochastic technique~\cite{Puccioni2015}, based on recurrence relations which make use of Einstein's semiclassical field theory~\cite{Einstein1917} (for more information cf. Supplementary Material available online).  Since the latter does not consider field phases, it is necessary to check the field's coherence in our experiment.  Transfer of these findings to other systems will entirely depend on the coherence length of the emission regime (low, for instance, in nanolasers in the threshold region) compared to the external cavity length.   

In the self-spiking regime ($i \lessapprox 1.1 i_{th}$), the emission resembles (filtered) Amplified Spontaneous Emission, thus one expects emission linewidths of the order of a nanometer (experimentally confirmed) and thus coherence lengths entirely negligible compared to the external feedback length $2 L_{ec}$.  In the partially coherent regime ($1.1 i_{th} \lessapprox i \lessapprox 2 i_{th}$), the spectral width is of the order of hundreds of $GHz$ (in the middle of the bias range), leading to coherence lengths $\mathcal{L}_c \approx K \, {\rm mm} \ll 2 L_{ec}$ ($K = O(1)$).  Thus, we can consider the feedback contribution as being constituted (mostly) of incoherent photons (as in~\cite{Ju2005}, but with an entirely different physical origin) reinjected into the cavity and simply added to the intracavity photon number.
This process is simulated by considering a stochastic transmission of a given fraction ($1.5\%$, as in the experimental estimates) of the backreflected photon number coming back through the output mirror, with a time delay equal to the experimentally measured one, and entering the cavity with a Poissonian probability law, in agreement with the whole conceptual framework of the Stochastic Simulator~\cite{Puccioni2015} (cf. Supplementary Material for additional information).

\begin{figure}[!t]
\centering
  \includegraphics[width=3.3in]{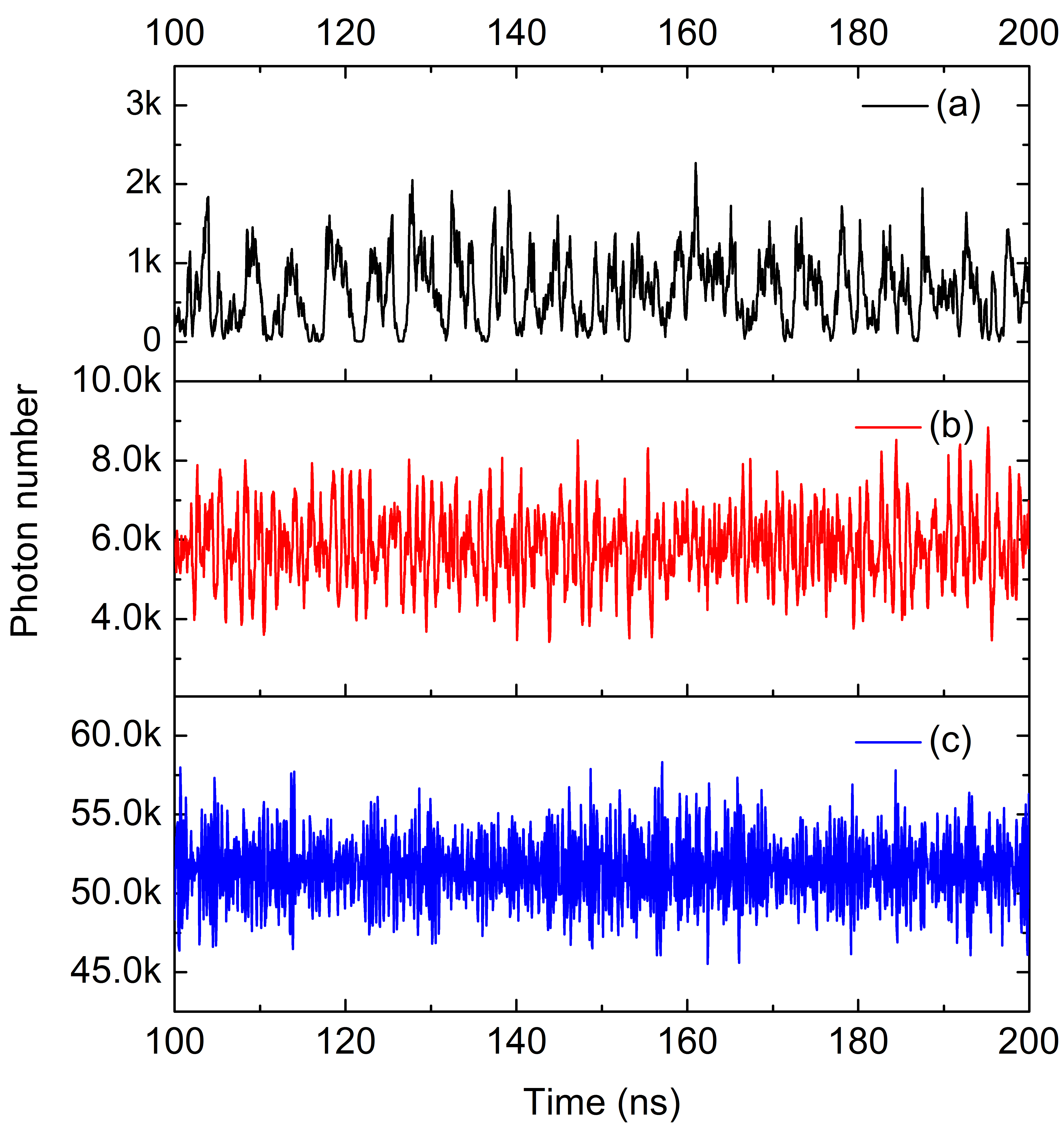}
  \caption{Numerical time traces computed for normalized pump values $\frac{P}{P_{th}}$: (a) $1.0$; (b) $1.1$, and (c) $2.0$.}
  \label{SDynamics}
\end{figure}

Fig.~\ref{SDynamics} shows a sample of the numerically predicted temporal laser dynamics in the presence of feedback for pump values corresponding to the experimental ones.  Comparison to the experimental measurements (Fig. \ref{EDynamics}) shows a good qualitative agreement and comparable features:  an irregular spiking behaviour is found in Fig.~\ref{SDynamics}a, while Fig.~\ref{SDynamics}b and Fig.~\ref{SDynamics}c show a noisier output.

\begin{figure}[!t]
\centering
  \includegraphics[width=3.2in]{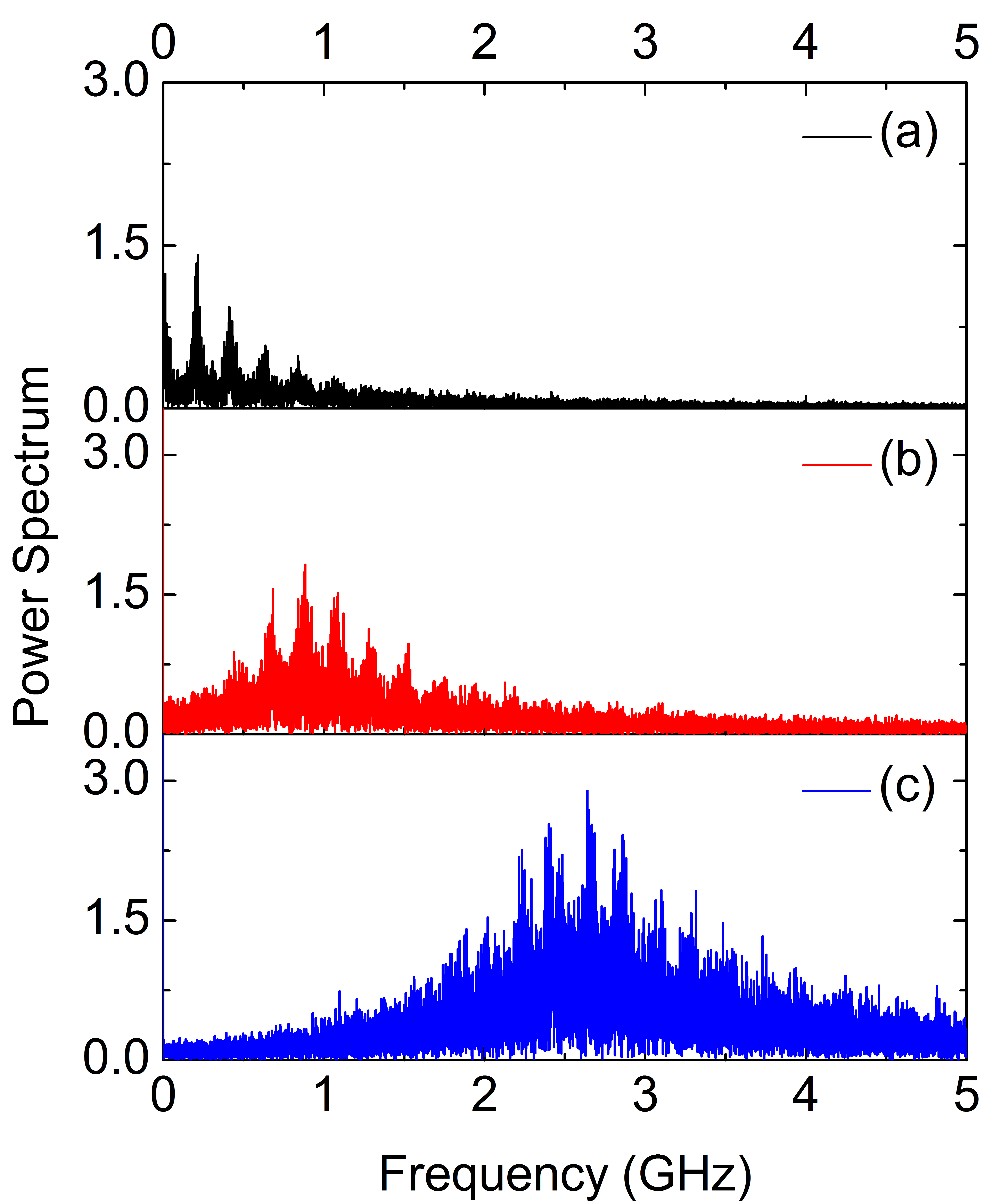}
  \caption{Numerical rf spectra for normalized pump values $\frac{P}{P_{th}}$: (a) $1.0$; (b) $1.1$, and (c) $2.0$.}
  \label{Spec}
\end{figure}

A good qualitative agreement also exists when comparing the numerically predicted rf spectra (Fig.~\ref{Spec}) with the experimentally measured ones:  in the self-spiking regime (panel a) there is no apparent background and the four peaks corresponding to the external cavity (multiples) are closely reproduced; at coherence onset (panel b) the intrinsic rf laser resonance appears with the superposed external cavity comb, while for established oscillation (but not full coherence! -- cf. Fig.~\ref{Exp-autoenlarge3}) the broad rf laser resonance moves between $2$ and $3 \, \rm GHz$ with the superposed frequency comb (Fig.~\ref{Spec}c).  The main discrepancy appears in the low-frequency part of the spectrum, which decreases instead of increasing (Fig.~\ref{ESpec}).  This is probably due to the fact that the model does not keep long-term memory and cannot properly reproduce the LFFs, in spite of a qualitative resemblance between the temporal traces (Figs.~\ref{EDynamics} and \ref{SDynamics}, panels (c)).

\begin{figure}[!t]
\centering
  \includegraphics[width=3.3in]{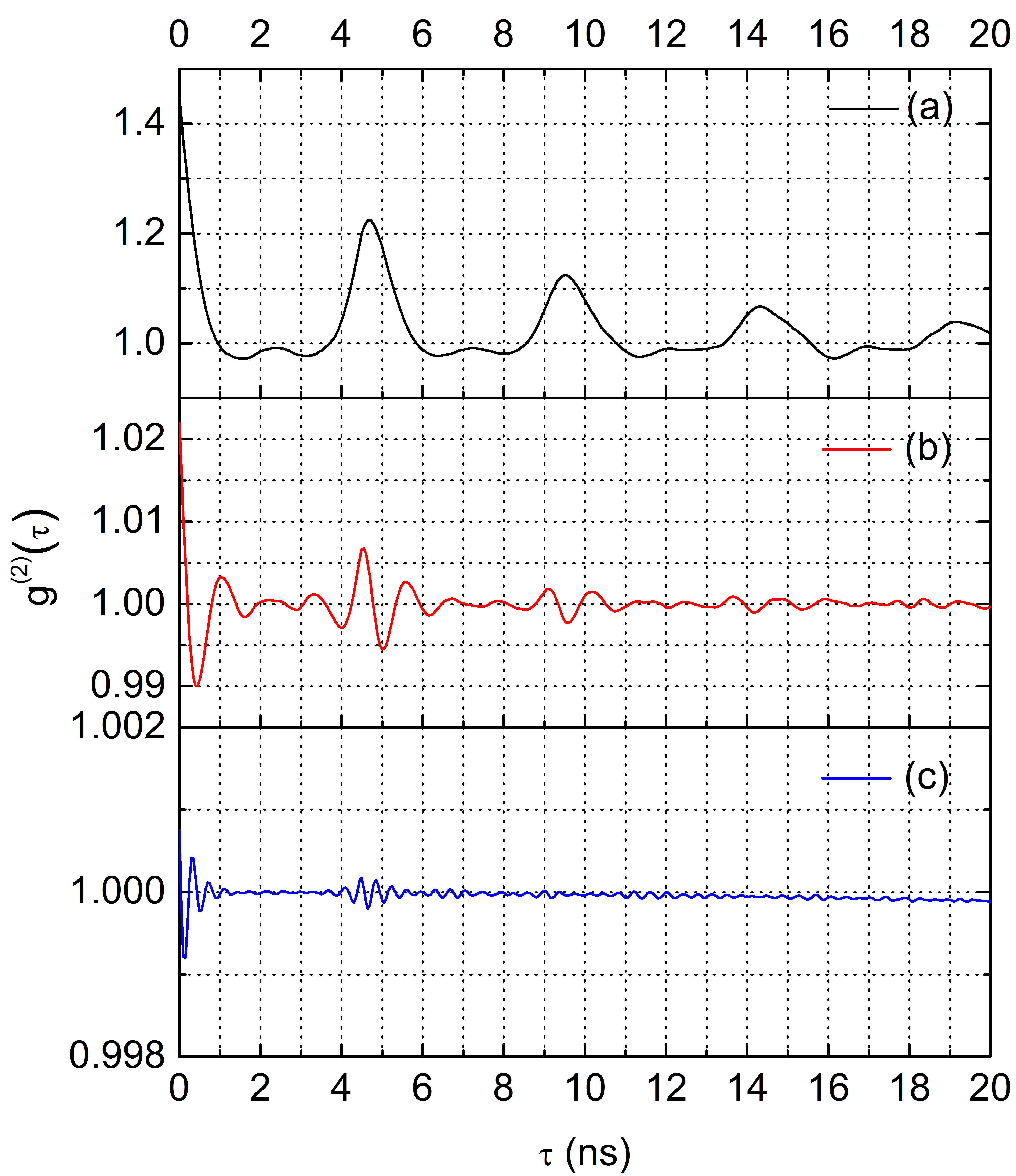}
  \caption{Numerical time-delayed second-order autocorrelation for normalized pump values $\frac{P}{P_{th}}$: (a) $1.0$; (b), $1.1$, and (c),$2.0$.}
  \label{SCorrelation}
\end{figure}

As a last step, we compare the predicted correlations, eq.~(\ref{defg2}), to the experimental ones in Fig.~\ref{SCorrelation}.  Aside from a somewhat smoother signal, devoid of the technical noise which affects the measurements (especially in the self-pulsing regime) the agreement is excellent!  All features analyzed in Fig.~\ref{ECorrelation} are found here, thus validating the interpretations that we have previously offered.  A careful look at Fig.~\ref{Spec}c shows a faster drop in the recurrences of the autocorrelation function, while the slightly negative overall slope is currently attributed to numerical problems.

\section{Conclusions}

An experiment performed in a micro-VCSEL, pumped in the transition region between incoherent and coherent emission with reinjection levels (from a ``long'' external cavity) compatible with parasitic reflections shows a lowering of the minimum bias value to obtain light emission accompanied by a reduction of the photon bursts typical of the free-running laser.  The influence of the optical reinjection remains moderate until the actual coherence threshold ($g^{(2)}(0) = 1$) is approached.

Comparison between traditional dynamical indicators and correlation functions shows that the latter are capable of characterizing the influence even of weak feedback, thus enabling them as a tool for the analysis of optical reinjection in nanolasers.  The possibility of basing the analysis of the laser dynamics on the sole second-order correlations represents a substantial step in furthering the investigation of the temporal behaviour of micro- and nanolasers even in the absence of sequential information.  Finally, we have proven that predictions obtained from fully stochastic numerical simulations, in the incoherent feedback regime, are in good qualitative agreement with the observations, except for the reproduction of low frequency fluctuations.  This is probably due to reliance of the numerical scheme on the photon number, while the field's phase progressively gains relevance in the gradual transition towards coherent laser emission.  Nonetheless, the scheme's validity in the fully incoherent regime promises to provide insights into the feedback dynamics at the smallest laser scales~\cite{Wang2018}.

\section*{Acknowledgment}
The authors are grateful to the R\'egion PACA and BBright for support, and to B. Garbin, F. Gustave and M. Marconi for assistance and discussions. Technical support from J.-C. Bery (mechanics) and from J.-C. Bernard and A. Dusaucy (electronics) is gratefully acknowledged. T. W. thanks the scientific research starting fund (KYS045618036) and national natural science foundation of China (61804036).  G.L. L. acknowledges discussions with T. Ackemann, M. Giudici, J. M\o rk and S. Reitzenstein.  The authors are grateful to two anonymous Referees for constructive criticism and valuable advice.

\ifCLASSOPTIONcaptionsoff
  \newpage
\fi



%

%

\begin{IEEEbiographynophoto}{Tao Wang}
received his Ph.D. degree in physics from the Universit\'e de Nice-Sophia Antipolis, France, in 2016. From 2013 to 2016, he worked in the Institut Non Lin\'eaire de Nice (now the Institut de Physique de Nice) as a PhD student. Since 2016, he was an post-doctoral fellow in the Institut National de la Recherche Scientifique, Canada. He currently is an associate professor in the School of Electronics and Information, Hangzhou Dianzi University, Hangzhou, China. His research interests include micro/nano scale laser dynamics, optical sensors based on lasers, nonlinear optics, light-matter interactions, and optical materials.
\end{IEEEbiographynophoto}

\begin{IEEEbiographynophoto}{Xianghu Wang}
is studying in Hangzhou Dianzi University, Hangzhou, China. His research interest includes semiconductor laser physics,  theoretical modelling and understanding of dynamics in micro-/nanoscale lasers
\end{IEEEbiographynophoto}

\begin{IEEEbiographynophoto}{Zhilei Deng}
is studying in Hangzhou Dianzi University, Hangzhou, China. His research interest focuses on theoretical design and modelling of small size VCSELs and optical sensors based on the laser devices.
\end{IEEEbiographynophoto}

\begin{IEEEbiographynophoto}{Jiacheng Sun}
is studying in Hangzhou Dianzi University, Hangzhou, China. His research interest mainly focuses on light-matter interaction in low dimensional structures, especially exploring novel optical properties of nanolasers.
\end{IEEEbiographynophoto}

\begin{IEEEbiographynophoto}{Gian Piero Puccioni}
earned his degree in Physics at the Universit\`a degli Studi di Firenze (Italy) and had a Post-doctoral Fellowship at the Univerity of Toronto (Canada). Between 1987 and 2004 he was a researcher at the National Institute of Optics in Florence where he also held the course in Numerical Computations of the Specialization School in Optics and was the Head of the Computer Center of the Institute. In 2004 he moved to the Institute of Complex Systems (CNR) in Firenze. He published several articles both experimental and theoretical on chaotic behavior in lasers, and more recently on mathematical models for nanolasers.
\end{IEEEbiographynophoto}


\begin{IEEEbiographynophoto}{Gaofeng Wang}
(S'93-M'95-SM'01) received the Ph.D. degree in electrical engineering from the University of Wisconsin-Milwaukee, Milwaukee, WI, USA, in 1993, and the Ph.D. degree in scientific computing from Stanford University, Stanford, CA, USA, in 2001. From 1993 to 1996, he was a Scientist with Tanner Research Inc., Pasadena, CA. From 1996 to 2001, he was a Principal Research and Development Engineer with Synopsys Inc., Mountain View, CA. In 1999, he served as a Consultant with Bell Laboratories, Murray Hill, NJ, USA. From 2001 to 2003, he was the Chief Technology Officer (CTO) of Intpax, Inc., San Jose, CA. From 2004 to 2010, he was the CTO of Siargo Inc., Santo Clara, CA. From 2004 to 2013, he was a Professor and the Head in the CJ Huang Information Technology Research Institute with Wuhan University, Wuhan, China. From 2010 to 2013, He was the Chief Scientist with Lorentz Solution, Inc., Santa Clara, CA. He is currently a Distinguished Professor with Hangzhou Dianzi University, Hangzhou. He has authored over 210 journal articles and holds 30 patents. His current research interests include integrated circuit and microelectromechanical system design and simulation, computational electromagnetics, electronic design automation, and wavelet applications in engineering.
\end{IEEEbiographynophoto}

\begin{IEEEbiographynophoto}{Gian Luca Lippi}
(Laurea in Fisica, University of Florence, Italy, 1984; Ph.D. Bryn Mawr College, USA, 1990; Habil. Dir. Rech. Universit\'e de Nice-Sophia Antipolis, France, 1998) is currently Distinguished Professor at the Physics Department of the Universit\'e C\^ote d'Azur and member of the Institut de Physique de Nice (formerly Institut Non Lin\'eaire de Nice) since 1994.  Between 1990 and 1993 he was active as Post-Doctoral Fellow at the Institut f\"ur Angewandte Physik (Westf\"{a}lische Wilhelms-Universit\"{a}t M\"{u}nster, Germany) first as Alexander-von-Humboldt Fellow, then with DFG support.  His research covers laser dynamics, nonlinear dynamics in optical systems, laser-matter interactions, particle trapping, and physical properties of nanolasers, mainly from an experimental point of view, but including modelling aspects.  He is co-author and author of over 60 papers in refereed journals, 24 conference proceedings and nearly 120 contributions to conferences (22 invited presentations).  He has been director of a French Doctoral School in Sciences and subsequently Director of the European Doctorate EDEMOM.  Former International Consultant for the project ``Laser Trapped Mirror Proposal'' (MSMT, NASA), past referee for INTAS (EU) and EPSRC (UK) programmes, referee for all the main physics journals, he is currently member of the Strategic Council of REA.
\end{IEEEbiographynophoto}




\end{document}